\documentstyle[12pt,graphicx,pstricks,braket,amsmath,amssymb,enumitem,
                    epstopdf,pdfpages,cite,tabu,hyperref]{article}
\numberwithin{equation}{section}
\newcommand{\ob}[1]{\bar{\textbf{#1}}}
\newcommand{\bb}[1]{{\textbf{#1}}}
\newcommand{\hi}[1]{}

\newcommand{\FbR}{{\bar F}_R}
\newcommand{\Db}{{\bar D}}
\newcommand{\Sb}{{\bar S}}
\newcommand{\phib}{{\bar \phi}}
\newcommand{\FL}{{F}_L}
\newcommand{\FR}{{F}_R}
\newcommand{\oh}{\frac{1}{2}}
\begin{document}

\def\AEF{A.E. Faraggi}

\def\JHEP#1#2#3{{JHEP} {\textbf #1}, (#2) #3}
\def\vol#1#2#3{{\bf {#1}} ({#2}) {#3}}
\def\NPB#1#2#3{{\it Nucl.\ Phys.}\/ {\bf B#1} (#2) #3}
\def\PLB#1#2#3{{\it Phys.\ Lett.}\/ {\bf B#1} (#2) #3}
\def\PRD#1#2#3{{\it Phys.\ Rev.}\/ {\bf D#1} (#2) #3}
\def\PRL#1#2#3{{\it Phys.\ Rev.\ Lett.}\/ {\bf #1} (#2) #3}
\def\PRT#1#2#3{{\it Phys.\ Rep.}\/ {\bf#1} (#2) #3}
\def\MODA#1#2#3{{\it Mod.\ Phys.\ Lett.}\/ {\bf A#1} (#2) #3}
\def\RMP#1#2#3{{\it Rev.\ Mod.\ Phys.}\/ {\bf #1} (#2) #3}
\def\IJMP#1#2#3{{\it Int.\ J.\ Mod.\ Phys.}\/ {\bf A#1} (#2) #3}
\def\nuvc#1#2#3{{\it Nuovo Cimento}\/ {\bf #1A} (#2) #3}
\def\RPP#1#2#3{{\it Rept.\ Prog.\ Phys.}\/ {\bf #1} (#2) #3}
\def\APJ#1#2#3{{\it Astrophys.\ J.}\/ {\bf #1} (#2) #3}
\def\APP#1#2#3{{\it Astropart.\ Phys.}\/ {\bf #1} (#2) #3}
\def\EJP#1#2#3{{\it Eur.\ Phys.\ Jour.}\/ {\bf C#1} (#2) #3}
\def\etal{{\it et al\/}}
\def\notE6{{$SO(10)\times U(1)_{\zeta}\not\subset E_6$}}
\def\E6{{$SO(10)\times U(1)_{\zeta}\subset E_6$}}
\def\highgg{{$SU(3)_C\times SU(2)_L \times SU(2)_R \times U(1)_C \times U(1)_{\zeta}$}}
\def\highSO10{{$SU(3)_C\times SU(2)_L \times SU(2)_R \times U(1)_C$}}
\def\lowgg{{$SU(3)_C\times SU(2)_L \times U(1)_Y \times U(1)_{Z^\prime}$}}
\def\SMgg{{$SU(3)_C\times SU(2)_L \times U(1)_Y$}}
\def\Uzprime{{$U(1)_{Z^\prime}$}}
\def\Uzeta{{$U(1)_{\zeta}$}}

\newcommand{\cc}[2]{c{#1\atopwithdelims[]#2}}
\newcommand{\bev}{\begin{verbatim}}
\newcommand{\beq}{\begin{equation}}
\newcommand{\ba}{\begin{eqnarray}}
\newcommand{\ea}{\end{eqnarray}}

\newcommand{\beqa}{\begin{eqnarray}}
\newcommand{\beqn}{\begin{eqnarray}}
\newcommand{\eeqn}{\end{eqnarray}}
\newcommand{\eeqa}{\end{eqnarray}}
\newcommand{\eeq}{\end{equation}}
\newcommand{\beqt}{\begin{equation*}}
\newcommand{\eeqt}{\end{equation*}}
\newcommand{\Eev}{\end{verbatim}}
\newcommand{\bec}{\begin{center}}
\newcommand{\eec}{\end{center}}
\newcommand{\bes}{\begin{split}}
\newcommand{\ees}{\end{split}}
\def\ie{{\it i.e.~}}
\def\eg{{\it e.g.~}}
\def\half{{\textstyle{1\over 2}}}
\def\nicefrac#1#2{\hbox{${#1\over #2}$}}
\def\third{{\textstyle {1\over3}}}
\def\quarter{{\textstyle {1\over4}}}
\def\m{{\tt -}}
\def\mass{M_{l^+ l^-}}
\def\p{{\tt +}}

\def\slash#1{#1\hskip-6pt/\hskip6pt}
\def\slk{\slash{k}}
\def\GeV{\,{\rm GeV}}
\def\TeV{\,{\rm TeV}}
\def\y{\,{\rm y}}

\def\l{\langle}
\def\r{\rangle}
\def\LRS{LRS  }

\begin{titlepage}
\samepage{
\setcounter{page}{1}
\rightline{LTH--1073}
%\rightline{arXiv:????.????}
\vspace{1.5cm}

\begin{center}
 {\Large \bf 
The $750$ GeV di--photon LHC excess and \\ \medskip 
Extra $Z^\prime$s in Heterotic--String Derived Models}
\end{center}

\begin{center}
%\vspace{1.cm}

%\vfill 
{\large
Alon E. Faraggi$^{\dagger}$\footnote{
		                  E-mail address: alon.faraggi@liv.ac.uk}
 and
John Rizos$^{\ddagger}$\footnote{ E-mail address: irizos@uoi.gr}}\\
\vspace{1cm}
{\it $^{\dagger}$ Dept.\ of Mathematical Sciences,
             University of Liverpool,
         Liverpool L69 7ZL, UK\\}
\vspace{.05in}
{\it $^{\ddagger}$ Department of Physics,
              University of Ioannina, GR45110 Ioannina, Greece\\}
\vspace{.025in}

\end{center}

\begin{abstract}

The ATLAS and CMS collaborations recently recorded possible di--photon 
excess at 750 GeV and a less significant di--boson excess 
around 1.9 TeV. Such excesses may be produced in heterotic--string
derived $Z^\prime$ models, where the di--photon 
excess may be connected with the Standard Model singlet scalar
responsible for the $Z^\prime$ symmetry breaking, whereas the 
di--boson excess arises from production of the 
extra vector boson. 
Additional vector--like states in the string $Z^\prime$ model are
instrumental to explain the relatively large width of 
the di--photon events and mandated by anomaly cancellation 
to be in the vicinity of the $Z^\prime$ breaking scale. 
Wilson line breaking of the non--Abelian gauge 
symmetries in the string models naturally gives 
rise to dark matter candidates. 
Future collider experiments 
will discriminate between the high--scale 
heterotic--string models, which preserve the 
perturbative unification paradigm indicated by the 
Standard Model data, versus the low scale string models.
We also discuss the possibility for the production of 
the diphoton events with high scale $U(1)_{Z^\prime}$
breaking.

\end{abstract}
\smallskip}
\end{titlepage}

\section{Introduction}

The Standard Model matter spectrum strongly favours its
embedding in $SO(10)$ chiral representations. This 
unification scenario is further supported by 
the perturbative logarithmic evolution of the 
Standard Model parameters; by proton longevity; 
and by the suppression of left--handed neutrino masses. 
This picture is reproduced in heterotic--string models 
\cite{heterotic, candelas}. 
The free fermionic formulation \cite{fff}
in particular has given rise 
to phenomenological three generation models that have 
been used to explore the unification of gravity and the 
gauge interactions. These models correspond to $Z_2\times Z_2$
orbifold compactification at special points in the Narain
moduli space and utilise discrete Wilson lines 
to break the non--Abelian gauge symmetry to an 
$SO(10)$ subgroup \cite{z2xz2}. The viable models constructed to date
include the flipped $SU(5)$ (FSU5) \cite{fsu5}; 
the standard--like models (SLM) \cite{slm};
the Pati--Salam models (PS) \cite{PSmodels}; 
and the left--right symmetric models (LRS) \cite{lrs};
whereas the $SU(4)\times SU(2)\times U(1)$ subgroup
did not produce viable models \cite{su421}. 
All these models give rise to extra observable gauge symmetries
at the string scale. Flavour universal symmetries typically arise
from the $SO(10)$ and $E_6$ group factors. However, 
preserving an unbroken extra gauge symmetry down to low scales
has proven to be elusive in the string models. The reasons being that
suppression of left--handed neutrino masses 
favours the breaking of lepton number at a high scale, 
whereas the $U(1)_\zeta$ gauge symmetry in the decomposition 
of $E_6\rightarrow SO(10)\times U(1)_\zeta$ tends to be 
anomalous in the string models and therefore cannot remain
unbroken down to low scales. 
This extra $U(1)$ 
in the string models is typically anomalous because 
of the projection of some states from the spectrum by the 
Generalised GSO (GGSO) projections, {\it i.e.} anomaly 
cancellation requires that the 
chiral spectrum appears in complete $E_6$ representations. 
However, the breaking of $E_6$ at the string scale mandates that 
the chiral states exist in incomplete $E_6$ multiplets.

Recently, however, we were able to construct a heterotic--string 
model in which the desired symmetry is anomaly free \cite{frzprime}. 
The derivation of this model utilises the spinor--vector duality
that was discovered in free fermionic heterotic--string models, 
and was obtained by using  
the classification methodology developed in
\cite{gkr, fknr,acfkr,frs}.
The model of ref. \cite{frzprime} 
is obtained from a self--dual $SO(10)$ model
under the exchange of the total number 
of spinorial $16\oplus\overline{16}$ and vectorial 
$10$ representations of $SO(10)$. This is the condition
that one has if the $SO(10)\times U(1)_\zeta$ symmetry
is enhanced to $E_6$. However, in the model of ref. 
\cite{frzprime} the
$SO(10)$ symmetry is not enhanced to $E_6$. 
This is the case in the free fermionic model 
if the different 16 and 10+1 representations,
that would form a complete 27 representation of $E_6$, 
are obtained from different fixed points of the underlying
$Z_2\times Z_2$ orbifold \cite{frzprime}. 
Adding the basis vector that breaks the $SO(10)$ symmetry
to the PS subgroup results in split multiplets, but the chiral 
spectrum still forms complete $E_6$ multiplets, hence rendering
$U(1)_\zeta$ anomaly free. We remark that while complete $E_6$ 
multiplets is sufficient for $U(1)_\zeta$ to be anomaly free, 
it might not be necessary and alternative possibilities may exist. 

The ATLAS and CMS collaborations \cite{atlas, cms} reported 
recently an excess in the di--photon searches at di--photon 
invariant mass of 750 GeV. This excess can be attributed to
a neutral scalar resonance of 750 GeV mass. 
Plausible candidates include 
the $SO(10)$ neutral singlet in the 27 of $E_6$, 
and $E_6$ singlets that arise in the string models, with 
the production and decay being produced via one--loop couplings
to heavy vector--like matter states. Such vector--like 
matter states are precisely those required from anomaly 
cancellation of the extra $Z^\prime$ gauge symmetry. 
Indications for an extra $Z^\prime$ of order 2 TeV have been earlier
suggested by the ATLAS and CMS experiments 
\cite{ATLASCMSzprime}. They have not been
substantiated by the run II data, but the possibility of an
extra $Z^\prime$ at that scale nevertheless remains. 
The signature of the string model with a low scale $Z^\prime$ in 
our model is characterised by the scalar resonance, 
by the extra $Z^\prime$ and 
by the additional vector--like matter
at the multi--TeV scale, required by anomaly cancellation. 

Alternative string constructions that may account for such
excesses have been recently suggested \cite{alternatives}, 
based on low scale string models and $F$--theory scenarios. 
Low scale string models give rise to Kaluza--Klein and heavy
string states. Therefore, future colliders will be able 
to discriminate between perturbative 
heterotic--string models and low scale string 
scenarios. Additionally, heterotic--string models 
give rise to states that do not satisfy the $E_6$ quantisation
of the $Z^\prime$ charges. Such states arise in string models
due to the breaking of the $E_6$ symmetry by Wilson lines and 
can produce viable dark matter candidates. This would be 
the case if the $Z^\prime$ symmetry is broken by a state 
that carries standard $E_6$ charge. A residual discrete 
symmetry then forbids the decay of the exotic string 
state to states that carry the standard $E_6$ charges. 
Such exotic states at the multi--TeV scale can provide 
viable thermal relics. On the other hand, if their exotic 
$Z^\prime$ charges can be determined experimentally, they 
provide a distinct signatures of the heterotic--string models 
that utilise discrete Wilson lines to break the $E_6$ 
symmetry. 

\section{The string model} 

The string model was constructed in ref. \cite{frzprime} and its
details will not be repeated here. The construction of the model
utilizes the free fermionic model building rules \cite{fff}, and
the notation that we use is prevalent in the literature 
(see {\it e.g.} \cite{fff, 
fsu5,slm, PSmodels, lrs, su421, frzprime} and references therein). 
The model is generated
by using the classification methods developed in \cite{gkr} for the 
classification of type IIB superstrings and extended in
\cite{fknr, acfkr, frs} for the classification of heterotic--string 
vacua with different $SO(10)$ subgroups.
The space of vacua is generated by working with a fixed 
set of boundary condition basis vectors and varying the 
GGSO coefficients \cite{fknr, acfkr, frs}, 
which are $\pm1$ phases in the one--loop partition function. 
The $Z^\prime$ model under consideration here was obtained in the class
of Pati--Salam heterotic string models, which are generated by a set of 
thirteen boundary condition basis vectors 
$
B=\{v_1,v_2,\dots,v_{13}\}. 
$
The basis vectors are shown in eq. 
(\ref{psbasis}),
\begin{eqnarray}
v_1=1&=&\{\psi^\mu,\
\chi^{1,\dots,6},y^{1,\dots,6}, \omega^{1,\dots,6}| \nonumber\\
& & ~~~\bar{y}^{1,\dots,6},\bar{\omega}^{1,\dots,6},
\bar{\eta}^{1,2,3},
\bar{\psi}^{1,\dots,5},\bar{\phi}^{1,\dots,8}\},\nonumber\\
v_2=S&=&\{\psi^\mu,\chi^{1,\dots,6}\},\nonumber\\
v_{2+i}=e_i&=&\{y^{i},\omega^{i}|\bar{y}^i,\bar{\omega}^i\}, \
i=1,\dots,6,\nonumber\\
v_{9}=z_1&=&\{\bar{\phi}^{1,\dots,4}\},\label{psbasis}\\
v_{10}=z_2&=&\{\bar{\phi}^{5,\dots,8}\},\nonumber\\
v_{11}=b_1&=&\{\chi^{34},\chi^{56},y^{34},y^{56}|\bar{y}^{34},
\bar{y}^{56},\bar{\eta}^1,\bar{\psi}^{1,\dots,5}\},\nonumber\\
v_{12}=b_2&=&\{\chi^{12},\chi^{56},y^{12},y^{56}|\bar{y}^{12},
\bar{y}^{56},\bar{\eta}^2,\bar{\psi}^{1,\dots,5}\},\nonumber\\
v_{13}=\alpha &=& \{\bar{\psi}^{4,5},\bar{\phi}^{1,2}\}. \nonumber
\end{eqnarray}
The fermions appearing in the curly brackets in eq. (\ref{psbasis})
are periodic, whereas those that do not appear are antiperiodic. 
The first twelve basis vectors in (\ref{psbasis}) generate the 
space of vacua with unbroken $SO(10)$ symmetry \cite{fknr}, whereas the 
thirteenth basis vector breaks the $SO(10)$ symmetry to the 
Pati--Salam subgroup \cite{acfkr}. 
The one--loop GGSO phases between the basis vectors are given by 
a $13\times 13$ matrix. Only the terms above the diagonal are
independent, whereas those on the diagonal and below are 
fixed by modular transformations \cite{fff}. Additional
constraints, such as requiring the vacuum to possess 
space--time supersymmetry, fixes additional phases and leaves 
a total of 66 independent phases. Using a random generation algorithm we 
can generate random choices of the independent phases. 
By imposing some physical criteria on the desired model, we 
can fish out models with desired physical characteristics. 
These include the absence of symmetry enhancing spacetime
vector bosons and exotic fractionally charged states from
the massless spectrum. A choice of GGSO phases that produces
these desired results is given by: 
\beq \label{BigMatrix}  (v_i|v_j)\ \ =\ \ \bordermatrix{
      & 1& S&e_1&e_2&e_3&e_4&e_5&e_6&b_1&b_2&z_1&z_2&\alpha\cr
 1    & 1& 1&  1&  1&  1&  1&  1&  1&  1&  1&  1&  1&      1\cr
S     & 1& 1&  1&  1&  1&  1&  1&  1&  1&  1&  1&  1&      1\cr
e_1   & 1& 1&  0&  0&  0&  0&  0&  0&  0&  0&  0&  0&      1\cr
e_2   & 1& 1&  0&  0&  0&  0&  0&  1&  0&  0&  0&  1&      0\cr
e_3   & 1& 1&  0&  0&  0&  1&  0&  0&  0&  0&  0&  1&      1\cr
e_4   & 1& 1&  0&  0&  1&  0&  0&  0&  0&  0&  1&  0&      0\cr
e_5   & 1& 1&  0&  0&  0&  0&  0&  1&  0&  0&  0&  1&      1\cr
e_6   & 1& 1&  0&  1&  0&  0&  1&  0&  0&  0&  1&  0&      0\cr
b_1   & 1& 0&  0&  0&  0&  0&  0&  0&  1&  1&  0&  0&      0\cr
b_2   & 1& 0&  0&  0&  0&  0&  0&  0&  1&  1&  0&  0&      1\cr
z_1   & 1& 1&  0&  0&  0&  1&  0&  1&  0&  0&  1&  1&      0\cr
z_2   & 1& 1&  0&  1&  1&  0&  1&  0&  0&  0&  1&  1&      0\cr
\alpha& 1& 1&  1&  0&  1&  0&  1&  0&  1&  0&  1&  0&      1\cr
  }
\eeq
In terms of the notation used in eq. (\ref{BigMatrix})
the GGSO one--loop coefficients in the $N=1$ partition function are 
given by
$$c{{vi}\atopwithdelims[] {v_j}}~=~{\rm exp}[i\pi(v_i|v_j)].$$
The full massless spectrum of the model, 
and its charges under the four dimensional gauge group, 
is given in ref \cite{frzprime}. 
%
%Here only the states in the observable twisted matter sector are shown in 
%table \ref{tableb}. 
In tables \ref{tableb} and \ref{tablehi} we provide a glossary of the 
states in the model and their charges under the 
$SU(4)\times SO(4)\times U(1)_\zeta$ gauge group.
We remark that we changed the notation of 
\cite{frzprime}
for the sextet fields from 
$D$ and ${\bar D}$ to $\Delta$ and ${\bar\Delta}$ 
to avoid confusion with the notation for the vector-like quarks below.
We note that the sextet states are in the vector representation 
of ${SO(6)}\equiv{SU(4)}$. 
They are vector--like with respect to the Standard Model subgroup,
but are chiral with respect to $U(1)_{\zeta}$. 
\begin{table}[!h]
\begin{tabular}{|c|c|c|c|}
\hline
Symbol& Fields in \cite{frzprime} & 
                         $SU(4)\times{SU(2)}_L\times{SU(2)}_R$&${U(1)}_{\zeta}$\\
\hline
$\FL$ &        $F_{1L},F_{2L},F_{3L}$&$\left({\bf4},{\bf2},{\bf1}\right)$&$+\oh$\\
$\FR$ &$F_{1R}$&$\left({\bf4},{\bf1},{\bf2}\right)$&$-\oh$\\
$\FbR$&${\bar F}_{1R},{\bar F}_{2R},{\bar F}_{3R},{\bar F}_{4R}$
                             &$\left({\bf\bar 4},{\bf1},{\bf2}\right)$&$+\oh$\\
$h$   &$h_1,h_2,h_3$&$\left({\bf1},{\bf2},{\bf2}\right)$&$-1$\\
$\Delta$&$ D_1,\dots,  D_7$&$\left({\bf6},{\bf1},{\bf1}\right)$&$-1$\\
$\bar\Delta$&$\Db_1,\Db_2,\Db_3,\Db_6$&$\left({\bf6},{\bf1},{\bf1}\right)$&$+1$\\
$S$&$\Phi_{12},\Phi_{13},\Phi_{23},\chi^+_1,\chi^+_2,\chi^+_3,\chi^+_5$
&$\left({\bf1},{\bf1},{\bf1}\right)$&$+2$\\
$\Sb$&$\bar\Phi_{12},\bar\Phi_{13},\bar\Phi_{23},\bar\chi^+_4$&$\left({\bf1},{\bf1},{\bf1}\right)$&$-2$\\
$\phi$&$\phi_1,\phi_2$&$\left({\bf1},{\bf1},{\bf1}\right)$&$+1$\\
$\phib$&$\bar\phi_1,\bar\phi_2$&$\left({\bf1},{\bf1},{\bf1}\right)$&$-1$\\
$\zeta$&$\Phi_{12}^-,\Phi_{13}^-,\Phi_{23}^-,\bar\Phi_{12}^-,\bar\Phi_{13}^-,\bar\Phi_{23}^-$&$\left({\bf1},{\bf1},{\bf1}\right)$&$\hphantom{+}0$\\
&$\chi_1^-,\chi_2^-,\chi_3^-,\bar\chi_4^-,\chi_5^-$&$$&$$\\
&$\zeta_i,\bar\zeta_i,i=1,\dots,9$&$$&$$\\
&$\Phi_i,i=1,\dots,6$&$$&$$\\
\hline
\end{tabular}
\caption{\label{tableb}
Observable sector field notation and associated states in \cite{frzprime}.}
\end{table}
\begin{table}[!h]
\begin{center}
\begin{tabular}{|c|c|c|c|}
\hline
Symbol& Fields in \cite{frzprime} & ${SU(2)}^4\times SO(8)$&${U(1)}_{\zeta}$\\
\hline
$H^+$&$H_{12}^3$&$\left({\bf2},{\bf2},{\bf1},{\bf1},{\bf1}\right)$&$+1$\\
&$H_{34}^2$&$\left({\bf1},{\bf1},{\bf2},{\bf2},{\bf1}\right)$&$+1$\\
$H^-$&$H_{12}^2$&$\left({\bf2},{\bf2},{\bf1},{\bf1},{\bf1}\right)$&$-1$\\
&$H_{34}^3$&$\left({\bf1},{\bf1},{\bf2},{\bf2},{\bf1}\right)$&$-1$\\
$H$&$H_{12}^1$&$\left({\bf2},{\bf2},{\bf1},{\bf1},{\bf1}\right)$&$0$\\
&$H_{13}^i,i=1,2,3$&$\left({\bf2},{\bf1},{\bf2},{\bf1},{\bf1}\right)$&$0$\\
&$H_{14}^i,i=1,2,3$&$\left({\bf2},{\bf1},{\bf1},{\bf2},{\bf1}\right)$&$0$\\
&$H_{23}^1$&$\left({\bf1},{\bf2},{\bf2},{\bf1},{\bf1}\right)$&$0$\\
&$H_{24}^1$&$\left({\bf1},{\bf2},{\bf1},{\bf2},{\bf1}\right)$&$0$\\
&$H_{34}^i,i=1,4,5$&$\left({\bf1},{\bf1},{\bf2},{\bf2},{\bf1}\right)$&$0$\\
$Z$&$Z_i,i=1,\dots,$&$\left({\bf1},{\bf1},{\bf8}\right)$&$0$\\
\hline
\end{tabular}
\end{center}
\caption{\label{tablehi}
Hidden sector field notation and associated states in \cite{frzprime}. }
\end{table}

The model is derived by fishing a self--dual model 
under the spinor--vector duality at the $SO(10)$ level, 
{\it i.e.} prior to incorporation of the basis vector $\alpha$. 
The $SO(10)$  model exhibits the self--duality property under the exchange
of the total number of spinorial plus anti--spinorial,
and the total number of vectorial $SO(10)$ representations. 
This is in fact a key ingredient in the construction of
the model and in the possibility of having an anomaly free 
$U(1)_\zeta$ as part of a low scale $Z^\prime$.
The spinor--vector duality was observed in the classification
of free fermionic vacua with $SO(10)$ GUT group 
\cite{svd1,svd2}. The statement is that for every vacuum 
with a total number \#1 of twisted ${16}\oplus\overline{16}$ spinorial
representations and a total number \#2 twisted $10$ vectorial
representations, there exist another vacuum in which the two 
are interchanged. To understand the origin of this duality 
it is instrumental to consider the case in which
$SO(10)\times U(1)_\zeta$ is enhanced to $E_6$. 
The chiral and anti--chiral representations
of $E_6$ are the $27$ and $\overline{27}$ representations 
that decompose under $SO(10)\times U(1)_\zeta$ as
$27=16_{1/2}+10_{-1}+1_2$
and 
$\overline{27}=\overline{16}_{-1/2}+10_{1}+1_{-2}$. 
Thus, in the case of $E_6$ the total number of ${16}\oplus\overline{16}$ 
is equal to the total number of $10$ representations, 
{\it i.e.} this case is self--dual under the spinor--vector
duality map. In the case of $E_6$, $U(1)_\zeta$ is anomaly free by
virtue of its embedding in $E_6$, whereas in vacua with broken 
$E_6$, $U(1)_\zeta$ is in general anomalous \cite{cleaveru1a}. 
Furthermore, the case of $E_6$ correspond to a 
string vacuum with $(2,2)$ worldsheet supersymmetry. The $N=2$
worldsheet supersymmetry on the bosonic side of the heterotic--string
has a spectral flow operator that exchanges between the spinorial
and vectorial components of the $E_6$ representations \cite{svd2}.
The vacua with broken $E_6$ symmetry only possess $(2,0)$ worldsheet
supersymmetry. In these vacua the would be spectral flow
operator induces the map between the spinor--vector dual vacua. 
The string vacua, however, also admit a class of self--dual vacua 
under the spinor--vector duality map without enhancement of the 
gauge symmetry to $E_6$. This is possible because
the different spinorial and vectorial components that make 
up complete $E_6$ representations are obtained from 
different fixed points of the underlying $Z_2\times Z_2$ 
orbifold. In this case the chiral spectrum still resides in
complete $E_6$ representations, and $U(1)_\zeta$ is anomaly free, 
but the gauge symmetry is not enhanced to $E_6$. 

The Pati--Salam heterotic string model generated by
eqs. (\ref{psbasis},\ref{BigMatrix}) breaks the
string matter states into the PS components, and as a result
some states are projected out. 
However, as can be seen from table \ref{tableb} the twisted
chiral matter spectrum of this model forms complete $E_6$ 
representations. It is noted that for $U(1)_\zeta$ to be anomaly
free only the chiral spectrum has to form complete 27 representations
of $E_6$, whereas the string model may contain vector--like 
states that do not form complete $E_6$ representations. 
The observable and hidden 
gauge symmetries at the string scale are generated by untwisted
states and are given by: 
\beqn
{\rm observable} ~: &~~~~~~~~SO(6)\times SO(4) \times 
U(1)_1 \times U(1)_2\times U(1)_3 \nonumber\\
{\rm hidden}     ~: &SO(4)^2\times SO(8)~~~~~~~~~~~~~~~~~~~~~~~~\nonumber
\eeqn
All the additional massless spacetime vector bosons that can enhance 
the observable and hidden gauge symmetries are projected 
out in this model due to the choice of GGSO phases in
eq. (\ref{BigMatrix}). The string model contains two anomalous
$U(1)$s with 
\beq
{\rm Tr}U(1)_1= 36 ~~~~~~~{\rm and}~~~~~~~{\rm Tr}U(1)_3= -36. 
\label{u1u3}
\eeq
Consequently, the $E_6$ combination
\beq
U(1)_\zeta= U(1)_1+ U(1)_2+ U(1)_3 
\label{u1zeta}
\eeq
is anomaly free and can be preserved as a component of an
extra $Z^\prime$ at lower scales. It should be emphasised that generically
$U(1)_\zeta$ is anomalous in the string models and therefore
has to  be broken in these models near the string scale \cite{dsw}.
It is anomaly free in the model generated by eqs. 
(\ref{psbasis},\ref{BigMatrix}) 
because the chiral spectrum forms complete 27 of $E_6$ multiplets.

%In addition to the states in table \ref{tableb} the string 
%model contains vector--like states from the untwisted and additional
%sectors. The untwisted sector gives rise the three pairs of 
%$SU(4)$ sextet fields denoted by $D_{1,2,3}$ and ${\bar D}_{1,2,3}$; 
%three pairs of $SO(10)$ singlets with $Q_{\zeta}=\pm2$ denoted by
%$\Phi_{12}$, ${\bar\Phi}_{12}$, 
%$\Phi_{13}$, ${\bar\Phi}_{13}$, 
%$\Phi_{23}$, ${\bar\Phi}_{23}$; 
%three pairs of $SO(10)$ singlets with $Q_{\zeta}=0$, which are 
%charged under the observable $U(1)_{1,2,3}$ gauge symmetries
%denoted by
%$\Phi_{12}^-$, ${\bar\Phi}_{12}^-$, 
%$\Phi_{13}^-$, ${\bar\Phi}_{13}^-$, 
%$\Phi_{23}^-$, ${\bar\Phi}_{23}^-$; 
%and six fields that are neutral under the entire 
%four dimensional gauge group denoted by
%$\Phi_i$, $i=\cdots6$. 
As seen from table \ref{tablehi} the model also contains 
vector--like states that transform under the hidden
sector $SU(2)^4\times SO(8)$ gauge group. They comprise four bidoublets  
denoted by $H^\pm$ that carry $Q_{\zeta}=\pm1$ charges, 
12 neutral bidoublets denoted by $H$ and five 
states that transform in the $8$ representation of the 
hidden $SO(8)$ gauge group with $Q_\zeta=0$.

In the notation of tables \ref{tableb} and \ref{tablehi}, 
the effective trilevel superpotential takes the form
\begin{eqnarray}
W= w+w'
\end{eqnarray} 
where
\begin{eqnarray}
w&= \FbR \FL h + \FbR \FbR \Delta+ \FL \FL \Delta + \FR \FR {\bar\Delta}+
\FR \FbR \zeta+ h h S 
\nonumber\\
& + 
\Delta \Delta S + \Db \Db \Sb+ \Delta \Db \zeta + \zeta \zeta \zeta + 
S \phib \phib+ \Sb \phi \phi
+ S \Sb \zeta\ ,\label{PSsupO}\\
w'&= S H^- H^+ + \Sb H^+ H^+ + S H H + \phi H^- H + \phib H^+ H
+\phi{\bar\phi}\zeta \nonumber\\
&+ \zeta H H + \zeta H^+ H^-+ Z Z \zeta \ , \label{PSsupH}
\end{eqnarray}
where we have suppressed all generation and field indices.
%It contains 16 fields
%that transform as bidoublets under the hidden sector $SU(2)^4$ 
%gauge group. 
%Of these four states carry $Q_{\zeta}=\pm1$ denoted by
%$H_{12}^2$, $H_{24}^2$, $H_{34}^3$ and $H_{12}^3$; 
%and twelve states carry $Q_{\zeta}=0$ denoted by
%$H_{12}^1$, $H_{34}^1$, $H_{34}^5$, $H_{14}^1$; 
%$H_{12}^2$, $H_{13}^1$, $H_{23}^1$, $H_{24}^1$; 
%$H_{13}^2$, $H_{14}^2$, $H_{13}^3$ and $H_{14}^3$.
%In the notation used here the lower subscript denotes the 
%two $SU(2)$ gauge groups under which the field transform 
%as doublets. 
%Additionally, the model contains five 
%states that transform in the $8$ representation of the 
%hidden $SO(8)$ gauge group with $Q_\zeta=0$. 
%The complete charges of the untwisted and hidden sector
%fields are given in \cite{frzprime}.  

As seen from table \ref{tableb} the string model contains the 
heavy Higgs states required to break the non--Abelian Pati--Salam
symmetry \cite{patisalam}. These are ${\cal H}=F_R$ and 
$\bar{\cal H}$, being a linear combination of the four 
$\bar{F}_R$ fields. The decomposition of these fields in terms of the 
Standard Model group factors is given by: 
\begin{align}
\bar{\cal H}({\bf\bar4},{\bf1},{\bf2})& \rightarrow u^c_H\left({\bf\bar3},
{\bf1},\frac 23\right)+d^c_H\left({\bf\bar 3},{\bf1},-\frac 13\right)+
                            {\bar {\cal N}}\left({\bf1},{\bf1},0\right)+
                             e^c_H\left({\bf1},{\bf1},-1\right)
                             \nonumber \\
{\cal H}\left({\bf4},{\bf1},{\bf2}\right) & 
\rightarrow  u_H\left({\bf3},{\bf1},-\frac 23\right)+
d_H\left({\bf3},{\bf1},\frac 13\right)+
              {\cal N}\left({\bf1},{\bf1},0\right)+ 
e_H\left({\bf1},{\bf1},1\right)\nonumber
\end{align}
The suppression of the left--handed neutrino masses favours
the breaking of the Pati--Salam symmetry at the high scale. 
Schematically, the neutrino seesaw mass matrix in terms of the 
Standard Model components takes the generic form \cite{PSmodels, tnm}
\begin{equation}
{\left(
\begin{matrix}
                 {\nu}, &{N}, &{\zeta}
\end{matrix}
   \right)}
  {\left(
\begin{matrix}
       0 &       (M_{_D})                &             0               \\
(M_{_D})  &          0                  & \langle{\bar {\cal N}}\rangle \\
       0   &\langle{\bar {\cal N}}\rangle &   \langle\zeta\rangle     \\
\end{matrix}
   \right)}
  {\left(
\begin{matrix}
                 {\nu}  \cr
                 {N}\cr
                 {\zeta} \cr
\end{matrix}
   \right)},
\label{nmm}
\end{equation}
where generation indices are suppressed, and  
$M_{_D}$ is the Dirac mass matrix arising from
the couplings of the chiral fermions to the light Higgs bi--doublets. 
The underlying $SO(10)$ symmetry dictates that the Dirac mass
term of the tau neutrino is proportional to 
that of the top quark \cite{tnm}. Hence, adequate suppression 
of the tau neutrino mass favours high scale breaking of $SU(2)_R$. 
More intricate scenarios in which $SU(2)_R$ is broken at a lower
scale, possibly even near the TeV scale, may be possible as well \cite{fg15}, 
however, for our purpose here we may assume that it is broken near 
the string scale. The breaking of the PS symmetry then leaves an unbroken 
$U(1)_{Z^\prime}$ symmetry given by
\beq
U(1)_{{Z}^\prime} ~=~
{3\over {10}} U(1)_{B-L} -{2\over5} U(1)_{T_{3_R}} - U(1)_\zeta
~\notin~ SO(10),
\label{uzpwuzeta}
\eeq
that may remain unbroken down to low scales provided that $U(1)_\zeta$ is
anomaly free. Furthermore, cancellation of the anomalies mandates 
the existence of additional vector--like quarks and leptons, arising
from the vectorial $10$ representation of $SO(10)$, as well as the $SO(10)$ 
singlet in the $27$ of $E_6$. The spectrum below the PS breaking
scale is displayed schematically 
in table \ref{table27rot}. The three right--handed 
neutrino $N_L^i$ fields are neutral under the gauge symmetry below
the $SU(2)_R$ breaking scale and are decoupled from the low 
energy spectrum. This condition is specific to the
extra $U(1)_{Z^\prime}$ combination in eq. (\ref{uzpwuzeta}). 
Here we assume that the spectrum is supersymmetric. 
We allowed for the possibility that the spectrum contains additional
pairs of vector--like electroweak Higgs doublets
and colour triplets. 
The spectrum may be compatible with GUT scale gauge coupling 
unification \cite{viraf}, where we may assume that the unification scale
is either at the GUT or string scales \cite{witten}, provided 
that there is an excess of one pair of vector--like electroweak doublets
beyond the number of pairs of vector--like colour triplets. 
This is possible in the free fermionic
heterotic--string models due to the doublet--triplet splitting mechanism 
that operates %on untwisted states 
in the string models \cite{dtsplitting}.
Additionally, we allowed for the possibility of light states
that are neutral under the 
$SU(3)_C\times SU(2)_L\times U(1)_Y\times U(1)_{Z^\prime}$ low
scale gauge group. 
The $U(1)_{Z^\prime}$ 
gauge symmetry can be broken at low scales by the VEV of the $SO(10)$ singlets
$S_i$ and/or ${\phi_{1,2}}$. 

\begin{table}[!h]
\noindent 
{\small
\begin{center}
{\tabulinesep=1.2mm
\begin{tabu}{|l|cc|c|c|c|}
\hline
Field &$\hphantom{\times}SU(3)_C$&$\times SU(2)_L $
&${U(1)}_{Y}$&${U(1)}_{Z^\prime}$  \\
%\hline
\hline
$Q_L^i$&    $3$       &  $2$ &  $+\frac{1}{6}$   & $-\frac{2}{5}$   ~~  \\
$u_L^i$&    ${\bar3}$ &  $1$ &  $-\frac{2}{3}$   & $-\frac{2}{5}$   ~~  \\
$d_L^i$&    ${\bar3}$ &  $1$ &  $+\frac{1}{3}$   & $-\frac{4}{5}$  ~~  \\
$e_L^i$&    $1$       &  $1$ &  $+1          $   & $-\frac{2}{5}$  ~~  \\
$L_L^i$&    $1$       &  $2$ &  $-\frac{1}{2}$   & $-\frac{4}{5}$  ~~  \\
%
%$N_L^i$&    $1$       &  $1$ &  ~~$0$            & ~~$0$   ~~  \\
\hline
$D^i$       & $3$     & $1$ & $-\frac{1}{3}$     & $+\frac{4}{5}$  ~~    \\
${\bar D}^i$& ${\bar3}$ & $1$ &  $+\frac{1}{3}$  &   $+\frac{6}{5}$  ~~    \\
$H^i$       & $1$       & $2$ &  $-\frac{1}{2}$   &  $+\frac{6}{5}$ ~~    \\
${\bar H}^i$& $1$       & $2$ &  $+\frac{1}{2}$   &   $+\frac{4}{5}$   ~~  \\
\hline
$S^i$       & $1$       & $1$ &  ~~$0$  &  $-2$       ~~   \\
\hline
$h$         & $1$       & $2$ &  $-\frac{1}{2}$  &  $-\frac{4}{5}$  ~~    \\
${\bar h}$  & $1$       & $2$ &  $+\frac{1}{2}$  &  $+\frac{4}{5}$  ~~    \\
\hline
${\cal D}$  & $3$       & $1$ &  $-\frac{1}{3}$  &  $+\frac{4}{5}$  ~~    \\
${\bar{\cal D}}$&${\bar 3}$& $1$ &  $+\frac{1}{3}$  &  $-\frac{4}{5}$  ~~ \\
\hline
$\phi$       & $1$       & $1$ &  ~~$0$         & $-1$     ~~   \\
$\bar\phi$       & $1$       & $1$ &  ~~$0$     & $+1$     ~~   \\
\hline
\hline
$\zeta^i$       & $1$       & $1$ &  ~~$0$  &  ~~$0$       ~~   \\
\hline
\end{tabu}}
\end{center}
}
\caption{\label{table27rot}
\it
Spectrum and
$SU(3)_C\times SU(2)_L\times U(1)_{Y}\times U(1)_{{Z}^\prime}$ 
quantum numbers, with $i=1,2,3$ for the three light 
generations. The charges are displayed in the 
normalisation used in free fermionic 
heterotic--string models. }
\end{table}

A distinct property of the free fermionic heterotic--string model, 
as noted from table \ref{tableb},
is the existence of the exotic states $\phi_{1,2}$, and $\bar\phi_{1,2}$. 
These states arise due to the breaking of $E_6$ by a discrete Wilson
line in the string model. Such states do not arise in pure field 
theory GUT models, and may be a distinct signature of the specific
string vacuum of eqs. (\ref{psbasis}, \ref{BigMatrix}), {\it i.e.}
they may be a distinct signature of the particular Wilson
line used in this model. These exotic states are $SO(10)$ singlets 
and therefore are neutral with respect to the Standard Model
gauge group. The breaking of the $U(1)_{Z^\prime}$ gauge symmetry
may leave a discrete symmetry that forbids the decay of these exotic 
states to the lighter Standard Model states. This is the 
case if the $U(1)_{Z^\prime}$ gauge symmetry is broken by one
of the $S_i$ states. The mass scale of the exotic states,
relative to the $Z^\prime$ breaking scale, 
then determines the type and 
whether they can provide a viable dark matter candidate \cite{ssr}.

\section{The di--photon events}\label{diphotons} 

The ATLAS \cite{atlas}
and CMS \cite{cms}
experiments reported in December 2015 an excess in the
the production  of di--photon events with a resonance around 750 GeV. It 
generated a flurry of activity with over 120 related papers since the 
announcement \cite{flurry}.
The statistical significance of the combined results is of the order
of 3 sigma. The more substantial indications are observed by the 
ATLAS experiment, which favours a rather broad width of the 
order of $\Gamma\sim45$ GeV, which is not incompatible with the
CMS results. The Landau--Yang theorem implies that only spin 0 or 2 resonance 
can decay into two photons. In the context of the perturbative 
heterotic--string construction the viable possibility is therefore
a spin 0 resonance decaying into two photons. The production 
can be generated via gluon fusion, similar to the signal that 
led to the discovery of the Standard Model Higgs boson via
$h\rightarrow 2\gamma$. There is no evidence at comparable 
energy scale of the resonance decaying into final states
with any other particles, {\it i.e.} with 
$t{\bar t}$,  
$b{\bar b}$, 
$\ell{\bar\ell}$, 
$ZZ$, $WW$, etc. A plausible explanation for the production and 
decay of such a resonance is via a Standard Model singlet scalar
field that couples to heavy vector--like quark and lepton pairs. 
In is noted that indeed scenarios along this lines have been 
proposed for such a resonance \cite{vectorlikeprops}.

\begin{figure}[t]
\begin{center}
\includegraphics[width=10cm]{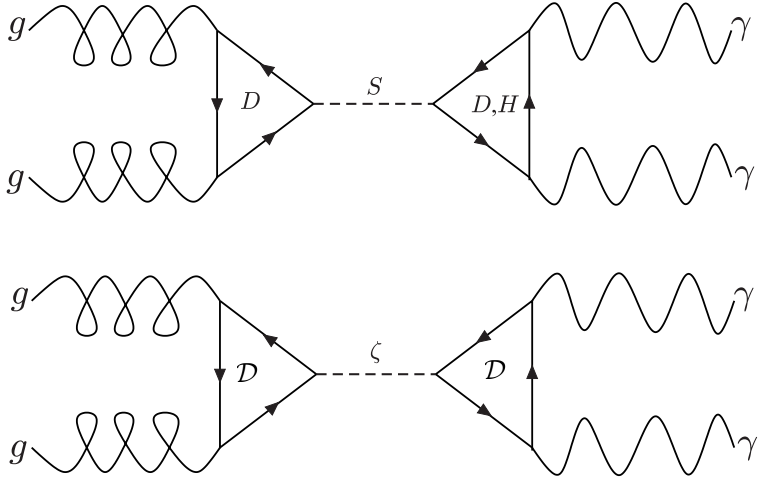}
\caption{Production and diphoton decay of the $SO(10)$ singlet state. 
The top diagram corresponds to production via the terms in 
eq. (\ref{suptermsS}) in the case with a low scale $U(1)_{Z^\prime}$, 
whereas the bottom diagram corresponds to production via the terms 
in eq. (\ref{suptermsZ}). 
}
\label{fig1}
\end{center}
\end{figure}

Turning to the low energy spectrum of the model in table \ref{table27rot}
it is noted that the $SO(10)$ singlets in the $27$ of $E_6$ $S_i$, 
as well as the $E_6$ singlets $\zeta_i$, provide
the needed fields. The low scale superpotential \ref{PSsupO}
gives rise to the terms 
\beqn
&
\lambda^D_{ijk}S^iD^j{\bar D}^k+
\lambda^H_{ijk}S_iH^j{\bar H}^k+
M^D_iD_i{\bar D}_i+
M_HH_i{\bar H}_i~, \label{suptermsS}\\
& \eta^{\cal D} \zeta {\cal D}  {\bar {\cal D}}+
\eta^h  \zeta h {\bar h}+
M_{\cal D} {\cal D}{\bar {\cal D}}+
M_hh{\bar h}~, 
\label{suptermsZ}
\eeqn
where we allowed for the possibility that the couplings arise from
terms that couple the vector--like states to the $S_i$ fields
that carry $Q_{Z^\prime}=-2$, as in eq. (\ref{suptermsS}), 
as well the coupling to the $\zeta_i$ fields that are 
neutral under $U(1)_{Z^\prime}$. 
These terms can generate the diphoton events via the diagram displayed 
in figure \ref{fig1}. Indeed, such terms are ubiquitous
in the string models. 
The cubic level and higher terms in the superpotential are 
calculated by using the tools developed in \cite{kln}. 
In the model of ref. \cite{frzprime} we find the 
terms for the states from table \ref{tableb} 
with couplings similar to those in eq. (\ref{suptermsS})
\begin{eqnarray}
\Delta \Delta S + h h S  
& = & 
~~{D_5}\,{D_7}\,{\chi_1^+}
+{D_3}\,{D_4}\,{\chi_1^+}
+{D_2}\,{D_5}\,{\chi_2^+}
+{D_4}\,{D_7}\,{\chi_2^+} 
+{D_2}\,{D_6}\,{\chi_3^+}
\nonumber\\
& &
+{D_1}\,{D_7}\,{\chi_5^+}
+{D_4}\,{D_5}\,{\chi_5^+}
+{D_1}\,{D_2}\,{\Phi_{12}}
+{D_2}\,{D_3}\,{\Phi_{23}}
+{D_1}\,{D_3}\,{\Phi_{13}}~~~~~~
\nonumber\\
& &
+{D_4}\,{D_4}\,{\Phi_{12}}
+{D_5}\,{D_5}\,{\Phi_{13}}
+{D_6}\,{D_6}\,{\Phi_{13}}
+{D_7}\,{D_7}\,{\Phi_{23}}
\nonumber\\
& &
+{h_2}\,{h_2}\,{\Phi_{13}}
+{h_3}\,{h_3}\,{\Phi_{13}}
+{h_1}\,{h_1}\,{\Phi_{12}}
+{h_1}\,{h_2}\,{\chi_5^+}
\label{so10supS}%\nonumber
\end{eqnarray}
as well as couplings similar to those in eq. (\ref{suptermsZ}) 
\begin{eqnarray}
\Delta \bar{\Delta} \zeta
&=&
+{\bar{D}_1}\,{D_7}\,{\chi_5^-}
+{\bar{D}_2}\,{D_5}\,{\chi_2^-}
+{\bar{D}_2}\,{D_6}\,{\chi_3^-}
+{\bar{D}_3}\,{D_4}\,{\chi_1^-}
\nonumber\\
& &
+{D_2}\,{\bar{D}_1}\,{\bar{\Phi}_{12}^-}
+{D_1}\,{\bar{D}_2}\,{\bar\Phi_{12}^-}
+{D_2}\,{\bar{D}_3}\,{\Phi_{23}^-}
+{D_3}\,{\bar{D}_2}\,{\bar{\Phi}_{23}^-}
\nonumber\\
& &
+{D_1}\,{\bar{D}_3}\,{\Phi_{13}^-}
+{D_3}\,{\bar{D}_1}\,{\bar{\Phi}_{13}^-}
+{D_1}\,{\bar{D}_6}\,{\bar{\chi}_4^-}
+{D_6}\,{\bar{D}_6}\,{\zeta_1}
\label{so10supZ}%\nonumber
\end{eqnarray}

%In the string model the fields denoted as 
%$\chi^+_{1, \dots, 5},~ \Phi_{12},~ \Phi_{23},~ \Phi_{13}$ 
%correspond to the $S_i$ fields, whereas fields denoted as
%$\chi^-_{1, \dots, 5}, ~ \Phi^-_{12},~ \Phi^-_{23},~ \Phi^-_{13}, 
%~ {\bar\Phi}^-_{12},~ {\bar\Phi}^-_{23},~ {\bar\Phi}^-_{13}, 
%\zeta_1$ correspond the $\zeta_i$ fields,  
%in table \ref{table27rot} and in figure \ref{fig1}. 
%The three Higgs bidoublets $h_{1,2,3}$ and the three $SU(4)$ sextets 
%$D_{4,5,6}$ can produce the corresponding
%low scale states $H, ~{\bar H}, ~D, ~{\bar D}$. The string model contains
%additional three pairs of vector--like sextet representations, which 
%arise from the untwisted sector \cite{frzprime}. 
The chiral 
spectrum of the model after $SU(2)_R$ breaking gives rise to three 
copies of the chiral states shown in table \ref{table27rot}. 
The string model does not give rise, however, to the extra pair
of vector--like Higgs doublets, which are instrumental for 
gauge coupling unification. The reason being that our 
string model uses symmetric boundary conditions rather 
than asymmetric boundary conditions \cite{dtsplitting}. 

The VEVs of the heavy Higgs fields ${\cal H}~,{\bar {\cal H}}$, 
%which can be identified
%with $F_{1R}$ and $\bar F_{1R}$ in table \ref{tableb} 
%leaves
the weak hypercharge combination given by
\beq
U(1)_Y=\frac{1}{2} U(1)_{B-L}+  U(1)_{T_{3_R}}
\label{u1y}
\eeq
unbroken, as well as the $U(1)_{Z^\prime}$ combination given
in eq. (\ref{uzpwuzeta}), which is orthogonal to the weak hypercharge 
combination. 
The scalar component of one of the $S_i$ fields provide the 
the Higgs field that breaks the extra $U(1)$ symmetry 
$\langle {\cal S}\rangle=v^\prime$. 
Provided that this VEV is of the order of the TeV scale, say $v'=5$ TeV, 
then ensures that the $U(1)_{Z^\prime}$ remains unbroken down to the TeV scale. 
Furthermore, ensuring that the extra $U(1)_{Z^\prime}$ is anomaly free
mandates that the extra vector--like quarks and leptons obtain their
mass of the order of $O(v')$ from the couplings in 
eq. (\ref{suptermsS}). 
Thus, all the ingredients needed to generate the characteristics 
of the diphoton events naturally exist in the string model. The model
then associates the diphoton events with the existence of additional
$U(1)_{Z^\prime}$ gauge symmetry at $O(v')$. With 
$M_{Z^\prime}\sim g^\prime v'$ and $M_{\cal S}\sim \lambda^\prime v^\prime$ 
being naively the masses of the heavy $Z^\prime$ vector boson and the 
Higgs field ${\cal S}$, respectively, we have that in this model the 
masses of the $Z^\prime$ vector field and the ${\cal S}$ scalar field
are closely related. These characteristics fit well with both a
di--boson excess at 1.9 TeV \cite{ATLASCMSzprime} as well as with the
di--photon excess at 750 GeV \cite{atlas, cms}. Furthermore, the model predicts
the existence of the additional vector--like quarks and leptons 
in the same vicinity. The existence of the $U(1)_{Z^\prime}$ symmetry
at the TeV scale, and the associated anomaly cancellation requirement, 
naturally explain the existence of the vector--like quarks and leptons at the 
$U(1)_{Z^\prime}$ breaking scale. However, as follows from 
eqs. (\ref{suptermsZ},\ref{so10supZ}), the diphoton events can also 
be mediated in the string models by scalars with $Q_{Z^\prime}=0$, 
and in this case the mass scale of the vector--like states 
is disassociated from the $U(1)_{Z^\prime}$ breaking scale. This is 
a less appealing scenario, but one which is allowed in the string 
model. In this case $U(1)_{Z^\prime}$ may be broken at a high scale 
along a flat direction by utilising the VEVs for, say ${\bar\chi}_4^+$
and $\chi_5^+$. 

\section{Conclusions}

Early indications from LHC run 1 and 2 provide evidence for excess in the 
di--photon channel with a resonance of the order of 750 GeV, 
and  a less significant di--boson excess at 1.9 Tev. Such signals
fit naturally in our heterotic--string derived model, with a high 
scale seesaw mechanism and a low scale $U(1)_{Z^\prime}$ breaking.
Furthermore, the model predicts the existence of additional 
vector--like quarks and leptons related to the $U(1)_{Z^\prime}$
breaking scale. 

%A key ingredient in the derivation of the heterotic--string model
%with an anomaly free $U(1)_\zeta$ is its self--duality property 
%under the spinor--vector duality discovered in the fermionic 
%$Z_2\times Z_2$ orbifolds. It has been suggested elsewhere that
%self--duality under the various string dualities \cite{sdvs}, 
%may play a role in the selection of the physical string vacuum. 
%The relevance of the self--duality under spinor vector
%duality to the derivation of the heterotic--string $U(1)_{Z^\prime}$ model
%provides another intriguing example in which self--duality plays an
%important role. 

Some of the proposals to explain the di--photon excess 
are in line with the explanation employed in our paper (see
{\it e.g.} \cite{vectorlikeprops}), and alternative scenarios 
have been proposed as well (see {\it e.g.} \cite{flurry}),
Some of these alternative scenarios employ a composite scalar 
singlet. The string derived model \cite{frzprime}
does contain hidden sector fields, charged under $U(1)_\zeta$, 
that can form composites that can mimic the charges of the 
singlet field $S_i$. Investigating whether this can provide an 
alternative scenario in the model at hand is left for future work.

The existence of vector--like quarks at the TeV scale 
poses a challenge when confronted with proton decay limits. 
Generically we anticipate that these states couple to the 
light quarks and may generate proton decay. How to 
avoid this conundrum remains a puzzle. Some plausible 
suggestions include the existence of local discrete symmetries 
that forbid the ominous couplings \cite{discrete}, 
and the special placement of matter fields in unified multiplets
\cite{fprt}.  

Alternative explanations have also been proposed in the case 
of models with a low string scale \cite{alternatives}.
The low scale string scenarios give rise to additional 
Kaluza--Klein and heavy string modes and therefore will 
be easily discerned from the heterotic--string scenarios.  
Explorations into the multi--TeV regime will adjudicate 
between the competing scenarios. We await the return of the 
collider.

\section*{Acknowledgments}

AEF thanks Shmuel Nussinov for discussions and the
theoretical physics groups at Oxford University, 
the Weizmann Institute and Tel Aviv University
for hospitality.
AEF is supported in part by the STFC (ST/L000431/1).
%JR research has been co-financed by the European
%Union (European Social Fund - ESF) and Greek national funds
%through the Operational Program
%“Education and Lifelong Learning" of the National Strategic 
%Reference Framework (NSRF) - Research Funding Program:
%THALIS Investing in the society of knowledge through the European
%Social Fund

\end{document}